\documentclass[11pt,onecolumn,draftcls]{IEEEtran}
\usepackage{amsmath,amssymb,cite,multirow}
\usepackage[dvips]{graphicx}
\usepackage{url}

\newtheorem{proposition}{Proposition}

\newtheorem{lemma}{Lemma}

\newtheorem{algorithm}{Algorithm}

\newcommand{\df}{\stackrel{\mbox{\scriptsize def}}{=}}

\newcommand{\rk}{\mathrm{rk}}

\newcommand{\Ac}{A_{\mbox{\tiny{C}}}}

\newcommand{\Bc}{B_{\mbox{\tiny{C}}}}

\newcommand{\dr}{d_{\mbox{\tiny{R}}}}
\newcommand{\ds}{d_{\mbox{\tiny{S}}}}

\newcommand{\di}{d_{\mbox{\tiny{I}}}}

\newcommand{\Ic}{I_{\mbox{\tiny{C}}}}

\newcommand{\Jc}{J_{\mbox{\tiny{C}}}}

\newcommand{\Kr}{K_{\mbox{\tiny{R}}}}

\newcommand{\Kc}{K_{\mbox{\tiny{C}}}}

\newcommand{\kc}{k_{\mbox{\tiny{C}}}}

\newcommand{\Nr}{N_{\mbox{\tiny{R}}}}

\newcommand{\Nc}{N_{\mbox{\tiny{C}}}}

\newcommand{\Vr}{V_{\mbox{\tiny{R}}}}

\newcommand{\Vc}{V_{\mbox{\tiny{C}}}}

\begin{document}
\title{Construction and Covering Properties of Constant-Dimension Codes}
\author{Maximilien Gadouleau and Zhiyuan Yan\\
Department of Electrical and Computer Engineering \\
Lehigh University, PA 18015, USA\\ E-mails: \{magc,
yan\}@lehigh.edu} \maketitle

\thispagestyle{empty}

\begin{abstract}
Constant-dimension codes (CDCs) have been investigated for
noncoherent error correction in random network coding. The maximum
cardinality of CDCs with given minimum distance and how to construct
optimal CDCs are both open problems, although CDCs obtained by
lifting Gabidulin codes, referred to as KK codes, are nearly
optimal. In this paper, we first construct a new class of CDCs based
on KK codes, referred to as augmented KK codes, whose cardinalities
are greater than previously proposed CDCs. We then propose a
low-complexity decoding algorithm for our augmented KK codes using
that for KK codes. Our decoding algorithm corrects more errors than
a bounded subspace distance decoder by taking advantage of the
structure of our augmented KK codes. In the rest of the paper we
investigate the covering properties of CDCs. We first derive bounds
on the minimum cardinality of a CDC with a given covering radius and
then determine the asymptotic behavior of this quantity. Moreover,
we show that liftings of rank metric codes have the highest possible
covering radius, and hence liftings of rank metric codes are \textbf{not
optimal} packing CDCs. Finally, we construct good covering CDCs by
permuting liftings of rank metric codes.
\end{abstract}
\IEEEpeerreviewmaketitle

\section{Introduction}\label{sec:introduction}
While random network coding \cite{ho_it06, FS_book07, HL08} has
proved to be a powerful tool for disseminating information in
networks, it is highly susceptible to errors caused by various
sources. Thus, error control for random network coding is critical
and has received growing attention recently. Error control schemes
proposed for random network coding assume two types of transmission
models: some (see, for example, \cite{cai_itw02,song_it03,yeung_cis06,cai_cis06,zhang_itw06,zhang_it08}) depend
on and take advantage of the underlying network topology or the
particular linear network coding operations performed at various
network nodes; others \cite{koetter_it08,silva_it08} assume that the
transmitter and receiver have no knowledge of such channel transfer
characteristics. The contrast is similar to that between coherent
and noncoherent communication systems. Data transmission in
noncoherent random network coding can be viewed as sending subspaces
through an operator channel \cite{koetter_it08}. Error correction in
noncoherent random network coding can hence be treated as a coding
problem where codewords are linear subspaces and codes are subsets
of the projective space of a vector space over a finite field.
Similar to codes defined over complex Grassmannians for noncoherent
multiple-antenna channels, codes defined in Grassmannians associated
with the vector space play a significant role in error control for
noncoherent random network coding; Such codes are referred to as
constant-dimension codes (CDCs) \cite{koetter_it08}. In addition to
the subspace metric defined in \cite{koetter_it08}, an injection
metric was defined for subspace codes over adversarial channels in
\cite{silva_arxiv08}.

Construction of CDCs has received growing attention in the
literature recently. In \cite{koetter_it08}, a Singleton bound for
CDCs and a family of codes were proposed, which are nearly
Singleton-bound-achieving and referred to as KK codes henceforth. A
multi-step construction of CDCs was proposed in
\cite{skachek_arxiv08}, and we call these codes Skachek codes;
Skachek codes have larger cardinalities than KK codes in some
scenarios, and reduce to KK codes otherwise. Further constructions
for small parameter values were given in \cite{kohnert_mmics08} and
the Johnson bound for CDCs was derived in \cite{xia_dcc09}. Although
the CDCs in \cite{xia_dcc09} are optimal in the sense of the Johnson
bound, they exist in some special cases only. Despite these previous
works, the maximum cardinality of a CDC with a given minimum
distance and how to construct optimal CDCs remain open problems.

Although the packing properties of CDCs were investigated in
\cite{kohnert_mmics08, xia_dcc09, skachek_arxiv08, koetter_it08},
the covering properties of CDCs have received little attention in
the literature. Covering properties are significant for error
control codes, and the covering radius is a basic geometric
parameter of a code \cite{cohen_book97}. For instance, the covering
radius can be viewed as a measure of performance: if the minimum
distance decoding is used, then the covering radius is the maximum
weight of a correctable error vector \cite{berger_book71}; if the
code is used for data compression, then the covering radius is a
measure of the maximum distortion \cite{berger_book71}. The covering
radius is also crucial for code design: if the covering radius is no
less than the minimum distance of a code, then there exists a
supercode with the same minimum distance and greater cardinality.

This paper has two main contributions. First, we introduce a new
class of CDCs, referred to as augmented KK codes.
The cardinalities of our augmented KK codes are
\textbf{always greater} than those of KK codes, and in \textbf{most} cases the
cardinalities of our augmented KK code are greater than those of
Skachek codes. Thus our augmented KK codes represent a step toward
solving the open problem (construction of optimal CDCs) mentioned above. Furthermore, we propose
an efficient decoding algorithm for our augmented KK codes using the
bounded subspace distance decoding algorithm in \cite{koetter_it08}.
Our decoding algorithm corrects more errors than a bounded subspace
distance decoder. Second, we investigate the covering properties of
CDCs. We first derive some key geometric results for Grassmannians.
Using these results, we derive upper and lower bounds on the minimum
cardinality of a CDC with a given covering radius. Since these
bounds are asymptotically tight, we also determine the asymptotic
behavior of the minimum cardinality of a CDC with a given covering
radius. Although liftings of rank metric codes can be used to
construct packing CDCs that are optimal up to a scalar (see, for example, those in \cite{koetter_it08}), we show that
all liftings of rank metric codes have the greatest covering radius
possible; our result further implies that liftings of rank metric codes are \textbf{not optimal} packing CDCs. We also
construct good covering CDCs by permuting liftings of rank metric
codes.

The rest of the paper is organized as follows. To be self-contained,
Section~\ref{sec:preliminaries} reviews some necessary background.
In Section~\ref{sec:construction}, we present our augmented KK codes
and a decoding algorithm for these codes. In
Section~\ref{sec:covering}, we investigate the covering properties
of CDCs.

\section{Preliminaries}\label{sec:preliminaries}

\subsection{Subspace codes}\label{sec:subspace_codes}
We refer to the set of all subspaces of $\mathrm{GF}(q)^n$ with
dimension $r$ as the Grassmannian of dimension $r$ and denote it as
$E_r(q,n)$; we refer to $E(q,n) = \bigcup_{r=0}^n E_r(q,n)$ as the
projective space. For $U,V \in E(q,n)$, both the \emph{subspace
metric} \cite[(3)]{koetter_it08}
\begin{equation} \label{eq:ds}
    \ds(U,V) \df \dim(U + V) - \dim(U \cap V) = 2\dim(U+V) - \dim(U) -
    \dim(V)
\end{equation}
and \emph{injection metric} \cite[Def.~1]{silva_arxiv08}
\begin{equation} \label{eq:di}
    \di(U,V) \df \frac{1}{2} \ds(U,V) + \frac{1}{2} |\dim(U) - \dim(V)|
    = \dim(U + V) - \min\{\dim(U), \dim(V)\}
\end{equation}
are metrics over $E(q,n)$. A {\em subspace code} is a nonempty
subset of $E(q,n)$. The minimum subspace (respectively, injection) distance of a subspace code is the
minimum subspace (respectively, injection) distance over all pairs
of distinct codewords.

\subsection{CDCs and rank metric codes}\label{sec:CDCs_and_rank_metric}

The Grassmannian $E_r(q,n)$ endowed with both the subspace and
injection metrics forms an association scheme \cite{koetter_it08,
delsarte_jct76}. For all $U,V \in E_r(q,n)$, $\ds(U,V) = 2\di(U,V)$
and the injection distance provides a natural distance spectrum,
i.e., $0\leq \di(U,V) \leq r$.
We have $|E_r(q,n)| = {n \brack r}$, where ${n \brack
r} = \prod_{i=0}^{r-1} \frac{q^n - q^i}{q^r - q^i}$ is the Gaussian
polynomial \cite{andrews_book76}, which satisfies
\begin{equation}
    q^{r(n-r)} \leq {n \brack r} < K_q^{-1} q^{r(n-r)}
    \label{eq:Gaussian}
\end{equation}
for all $0 \leq r \leq n$, where $K_q = \prod_{j=1}^\infty
(1-q^{-j})$ \cite{gadouleau_it08_dep}. We denote the number of
subspaces in $E_r(q,n)$ at distance $d$ from a given subspace as
$\Nc(d) = q^{d^2} {r \brack d} {n-r \brack d}$ \cite{koetter_it08},
and denote a ball in $E_r(q,n)$ of radius $t$ around a subspace $U$
and its volume as $B_t(U)$ and $\Vc(t) = \sum_{d=0}^t \Nc(d)$,
respectively.

A subset of $E_r(q,n)$ is called a constant-dimension code (CDC). A
CDC is thus a subspace code whose codewords have the same dimension.
We denote the \textbf{maximum} cardinality of a CDC in $E_r(q,n)$
with minimum distance $d$ as $\Ac(q,n,r,d)$. Constructions of CDCs
and bounds on $\Ac(q,n,r,d)$ have been given in \cite{koetter_it08,
xia_dcc09, skachek_arxiv08, gabidulin_isit08, kohnert_mmics08}. In
particular, $\Ac(q,n,r,1) = {n \brack r}$ and it is shown
\cite{skachek_arxiv08, xia_dcc09} for $r \leq \left\lfloor
\frac{n}{2}\right\rfloor$ and $2 \leq d \leq r$,
\begin{equation}\label{eq:bounds_Ac}
    \frac{q^{n(r-d+1)} - q^{(r+l)(r-d+1)}}{q^{r(r-d+1)} - 1}
    \leq \Ac(q,n,r,d) \leq
    \frac{{n \brack r-d+1}}{{r \brack r-d+1}},
\end{equation}
where $l \equiv n \mod r$. We denote the lower bound on
$\Ac(q,n,r,d)$ in (\ref{eq:bounds_Ac}) as $L(q,n,r,d)$. Since the
lower bound is due to the class of codes proposed by Skachek
\cite{skachek_arxiv08}, we refer to these codes as Skachek codes.

CDCs are closely related to rank metric codes \cite{delsarte_jct78,
gabidulin_pit0185, roth_it91}, which can be viewed as sets of
matrices in $\mathrm{GF}(q)^{m \times n}$. The rank distance between
two matrices ${\bf C}, {\bf D} \in \mathrm{GF}(q)^{m \times n}$ is
defined as $\dr({\bf C}, {\bf D}) \df \rk({\bf C} - {\bf D})$. The
maximum cardinality of a rank metric code in $\mathrm{GF}(q)^{m
\times n}$ with minimum distance $d$ is given by
$\min\{q^{m(n-d+1)}, q^{n(m-d+1)}\}$ and codes that achieve this
cardinality are referred to as MRD codes. In this paper, we shall
only consider MRD codes that are either introduced independently in
\cite{delsarte_jct78, gabidulin_pit0185, roth_it91} for $n \leq m$,
or their transpose codes for $n > m$. The number of matrices in
$\mathrm{GF}(q)^{m \times n}$ with rank $d$ is denoted as
$\Nr(q,m,n,d) = {n \brack d} \prod_{i=0}^{d-1} (q^m - q^i)$, and the
volume of a ball with rank radius $t$ in $\mathrm{GF}(q)^{m \times
n}$ as $\Vr(q,m,n,t) = \sum_{d=0}^t \Nr(q,m,n,d)$. The minimum
cardinality $\Kr(q^m,n,\rho)$ of a code in $\mathrm{GF}(q)^{m \times
n}$ with rank covering radius $\rho$ is studied in
\cite{gadouleau_it08_covering, gadouleau_cl08} and satisfies
$\Kr(q^m,n,\rho) = \Kr(q^n,m,\rho)$ \cite{gadouleau_it08_covering}.

CDCs are related to rank metric codes through the lifting operation
\cite{silva_it08}. Denoting the row space of a matrix ${\bf
M}$ as $R({\bf
M})$, the lifting of ${\bf C} \in \mathrm{GF}(q)^{r \times (n-r)}$
is defined as $I({\bf C}) = R({\bf I}_r | {\bf C}) \in E_r(q,n)$.
For all ${\bf C}, {\bf D} \in \mathrm{GF}(q)^{r \times (n-r)}$, we
have $\di(I({\bf C}), I({\bf D})) = \dr({\bf C}, {\bf D})$
\cite{silva_it08}. A KK code in $E_r(q,n)$ with minimum injection
distance $d$ is the lifting $I(\mathcal{C}) \subseteq E_r(q,n)$ of
an MRD code $\mathcal{C} \subseteq \mathrm{GF}(q)^{ r \times (n-r)}$
with minimum rank distance $d$ and cardinality
$\min\{q^{(n-r)(r-d+1)}, q^{r(n-r-d+1)} \}$. An efficient bounded
subspace distance decoding algorithm for KK codes was also given in
\cite{koetter_it08}. Although the algorithm was presented for $r
\leq \frac{n}{2}$, it can be easily generalized to all $r$.

\section{Construction of CDCs} \label{sec:construction}
In this section, we construct a new class of CDCs which contain KK
codes as proper subsets. Thus we call them augmented KK codes. We
will show that the cardinalities of our augmented KK codes are
always greater than those of KK codes, and that in most cases the
cardinalities of our augmented KK code are greater than those of
Skachek codes. Furthermore, we propose a low-complexity decoder for
our augmented KK codes based on the bounded subspace distance
decoder in \cite{koetter_it08}. Since dual CDCs preserve the
distance, we assume $r \leq \frac{n}{2}$ without loss of generality.

\subsection{Augmented KK codes}
Our augmented KK code is so named because it has a layered structure
and the first layer is simply a KK code. We denote a KK code in
$E_r(q,n)$ with minimum injection distance $d$ ($d\leq r$ by
definition) and cardinality $q^{(n-r)(r-d+1)}$ as $\mathcal{E}^0$.
For $1 \leq k \leq \left\lfloor \frac{r}{d} \right\rfloor$, we first
define two MRD codes $\mathcal{C}^k$ and $\mathcal{D}^k$, and then
construct $\mathcal{E}^k$ based on $\mathcal{C}^k$ and
$\mathcal{D}^k$. $\mathcal{C}^k$ is an MRD code in
$\mathrm{GF}(q)^{(r-kd) \times kd}$ with minimum distance $d$ for $k
\leq \left\lfloor \frac{r}{d} \right\rfloor - 1$ ($\left\lfloor
\frac{n-r}{d} \right\rfloor \geq \left\lfloor \frac{r}{d}
\right\rfloor$) and $\mathcal{C}^{\left\lfloor \frac{r}{d}
\right\rfloor} = \{ {\bf 0} \} \subseteq \mathrm{GF}(q)^{\left( r -
\left\lfloor \frac{r}{d} \right\rfloor d \right) \times \left\lfloor
\frac{r}{d} \right\rfloor d}$; $\mathcal{D}^k$ is an MRD code in
$\mathrm{GF}(q)^{r \times (n-r-kd)}$ with minimum distance $d$ for
$k \leq \left\lfloor \frac{n-r}{d} \right\rfloor - 1$ and
$\mathcal{D}^{\left\lfloor \frac{n-r}{d} \right\rfloor} = \{ {\bf
0}\} \subseteq \mathrm{GF}(q)^{r \times \left( n-r- \left\lfloor
\frac{n-r}{d} \right\rfloor d \right)}$. For $1 \leq k <
\left\lfloor \frac{r}{d} \right\rfloor$, the block lengths of
$\mathcal{C}^k$ and $\mathcal{D}^k$ are at least $d$, and hence
existence of MRD codes with the parameters mentioned above is
trivial. For $1 \leq k \leq \left\lfloor \frac{r}{d} \right\rfloor$,
$I(\mathcal{C}^k)$ and $I(\mathcal{D}^k)$ are either trivial codes
or KK codes with minimum injection distance $d$ in $E_{r-kd}(q,r)$
and $E_r(q,n-kd)$, respectively. For $1 \leq k \leq \left\lfloor
\frac{r}{d} \right\rfloor$, ${\bf C}_i^k \in \mathcal{C}^k$, and
${\bf D}_j^k \in \mathcal{D}^k$, we define $E_{i,j}^k \in E_r(q,n)$
as the row space of $\left(
\begin{array}{c|c|c|c}
    {\bf I}_{r-kd}
    & {\bf C}_i^k & {\bf 0} & \multirow{2}{*}{${\bf D}_j^k$}\\
    \cline{1-3}
    {\bf 0} & {\bf 0} & {\bf I}_{kd} &
\end{array} \right)$ and $\mathcal{E}^k = \{E_{i,j}^k\}_{i,j=0}^{|\mathcal{C}^k|-1,|\mathcal{D}^k|-1}$.
Our augmented KK code is simply $\mathcal{E} =
\bigcup_{k=0}^{\left\lfloor \frac{r}{d} \right\rfloor}
\mathcal{E}^k$. In order to determine its minimum distance, we first
establish two technical results. First, for any two matrices ${\bf
A} \in \mathrm{GF}(q)^{a \times n}$, ${\bf B} \in \mathrm{GF}(q)^{b
\times n}$, by (\ref{eq:ds}) and (\ref{eq:di}) we can easily show
that
\begin{eqnarray}
    \label{eq:ds_bound}
    \ds(R({\bf A}), R({\bf B})) &=& 2\rk({\bf A}^T | {\bf B}^T) -
    \rk({\bf A}) - \rk({\bf B}) \geq |\rk({\bf A}) - \rk({\bf B})|,\\
    \label{eq:di_bound}
    \di(R({\bf A}), R({\bf B})) &=& \rk({\bf A}^T | {\bf B}^T) -
    \min\{\rk({\bf A}), \rk({\bf B})\} \geq |\rk({\bf A}) - \rk({\bf B})|.
\end{eqnarray}
Second, we show that truncating the generator
matrices of two subspaces in $E(q,n)$ can only reduce the (subspace
or injection) distance between them.
\begin{lemma} \label{lemma:truncate}
Suppose $0 \leq n_1 \leq n$. Let ${\bf A} = ({\bf A}_ 1| {\bf A}_2)
\in \mathrm{GF}(q)^{a \times n}$, ${\bf B} = ({\bf B}_ 1| {\bf B}_2)
\in \mathrm{GF}(q)^{b \times n}$, where ${\bf A}_1 \in
\mathrm{GF}(q)^{a \times n_1}$ and ${\bf B}_1 \in \mathrm{GF}(q)^{b
\times n_1}$. Then for $i=1$ and $2$, $\ds(R({\bf A}_i), R({\bf
B}_i)) \leq \ds(R({\bf A}), R({\bf B}))$ and $\di(R({\bf A}_i),
R({\bf B}_i)) \leq \di(R({\bf A}), R({\bf B}))$.
\end{lemma}

\begin{proof}
It suffices to prove it for $i=1$ and $n_1 = n-1$. We need to
distinguish two cases, depending on $\rk({\bf A}_1^T | {\bf
B}_1^T)$. First, if $\rk({\bf A}_1^T | {\bf B}_1^T) = \rk({\bf A}^T
| {\bf B}^T)$, then it is easily shown that $\rk({\bf A}_1) =
\rk({\bf A})$ and $\rk({\bf B}_1) = \rk({\bf B})$, and hence
$\ds(R({\bf A}_1), R({\bf B}_1)) = \ds(R({\bf A}), R({\bf B}))$ and
$\di(R({\bf A}_1), R({\bf B}_1)) = \di(R({\bf A}), R({\bf B}))$ by
(\ref{eq:ds_bound}) and (\ref{eq:di_bound}), respectively. Second,
if $\rk({\bf A}_1^T | {\bf B}_1^T) = \rk({\bf A}^T | {\bf B}^T) -
1$, then $\ds(R({\bf A}_1), R({\bf B}_1)) = 2 \rk({\bf A}^T | {\bf
B}^T) - 2 - \rk({\bf A}_1) - \rk({\bf B}_1) \leq \ds(R({\bf A}),
R({\bf B}))$ by (\ref{eq:ds_bound}) and $\di(R({\bf A}_1), R({\bf
B}_1)) = \rk({\bf A}^T | {\bf B}^T) - 1 - \min\{\rk({\bf A}_1),
\rk({\bf B}_1)\} \leq \di(R({\bf A}), R({\bf B}))$ by
(\ref{eq:di_bound}).
\end{proof}

\begin{proposition} \label{prop:E_union_K}
$\mathcal{E}$ has minimum injection distance $d$.
\end{proposition}

\begin{proof}
We show that any two codewords $E_{i,j}^k, E_{a,b}^c \in
\mathcal{E}$ are at injection distance at least $d$ using
Lemma~\ref{lemma:truncate}. When $c \neq k$, let us assume $c < k$
without loss of generality, and then $\di(E_{i,j}^k, E_{a,b}^c) \geq
\di(R({\bf I}_{r-kd}| {\bf 0}), R({\bf I}_{r-cd})) = (k-c)d \geq d$.
When $c = k$ and $a \neq i$, then $\di(E_{i,j}^k, E_{a,b}^k) \geq
\di(I({\bf C}_i^k), I({\bf C}_a^k)) \geq d$. When $c = k$, $a = i$,
and $b \neq j$, then $\di(E_{i,j}^k, E_{i,b}^k) \geq \di(I({\bf
D}_j^k), I({\bf D}_b^k)) \geq d$.
\end{proof}

Let us first determine the cardinality of our augmented KK codes. By
construction, $\mathcal{E}$ has cardinality $|\mathcal{E}| =
q^{(n-r)(r-d+1)} + \sum_{k=1}^{\left\lfloor \frac{r}{d}
\right\rfloor} |\mathcal{C}^k| |\mathcal{D}^k|$, where
$|\mathcal{C}^{\left\lfloor \frac{r}{d} \right\rfloor}| = 1$ and
$|\mathcal{C}^k| = \min\{ q^{(r-kd)(kd-d+1)}, q^{kd(r-kd-d+1)} \}$
for $1 \leq k \leq \left\lfloor \frac{r}{d} \right\rfloor - 1$ and
$|\mathcal{D}^{\left\lfloor \frac{n-r}{d} \right\rfloor}| = 1$ and
\\ $|\mathcal{D}^k| = \min\{ q^{r(n-r-kd-d+1)}, q^{(n-r-kd)(r-d+1)}
\}$ for $1 \leq k \leq \left\lfloor \frac{n-r}{d} \right\rfloor -
1$.

Let us compare the cardinality of our augmented KK codes to those of
KK and Skachek codes. Note that all three codes are CDCs with minimum injection
distance $d$ in $E_r(q,n)$. First, it is easily shown that our augmented KK codes properly
contain KK codes for all parameter values. This is a clear
distinction from Skachek codes with cardinality $L(q,n,r,d)$, which
by (\ref{eq:bounds_Ac}) reduce to KK codes for $3r > n$. In order to
compare our codes to Skachek codes when $3r \leq n$, we first remark
that (\ref{eq:bounds_Ac}) and (\ref{eq:Gaussian}) lead to
$L(q,n,r,d) - q^{(n-r)(r-d+1)} < K_q^{-1} q^{(n-2r)(r-d+1)}$. Also,
we have $|\mathcal{E}| \geq q^{(n-r)(r-d+1)} +
|\mathcal{C}^1||\mathcal{D}^1| \geq q^{(n-r)(r-d+1)} +
q^{(n-r-d)(r-d+1)}$. Hence $|\mathcal{E}| - q^{(n-r)(r-d+1)} > K_q
q^{(r-d)(r-d+1)} (L(q,n,r,d) - q^{(n-r)(r-d+1)})$, and our augmented
KK codes have a greater cardinality than Skachek codes when $d < r$. We emphasize that
for CDCs of dimension $r$, their minimum injection distance $d$
satisfies $d\leq r$.
A Skachek code is constructed in multiple steps, and in the $i$-th
step ($i\geq 1$), subspaces that correspond to a KK code in
$E_r(q,n-ir)$ are added to the code. When $d=r$, $\mathcal{E}$ is
actually the code obtained after the first step.

\subsection{Decoding of augmented KK codes}
Let ${\bf A} = ({\bf A}_0 | {\bf A}_3) \in \mathrm{GF}(q)^{a \times
n}$ be the received matrix, where ${\bf A}_0 \in \mathrm{GF}(q)^{a
\times r}$ and ${\bf A}_3 \in \mathrm{GF}(q)^{a \times (n-r)}$.
We propose a decoding algorithm that either produces the unique
codeword in $\mathcal{E}$ closest to $R({\bf A})$ in the subspace
metric or returns a failure. Suppose the minimum subspace
distance of our augmented KK codes is denoted as $2d$, a bounded
distance decoder would find the codeword that is closest to $R({\bf A})$ up to subspace distance $d-1$. Our decoding algorithm always
returns the correct codeword if it is at subspace distance at most
$d-1$ from the received subspace, thus correcting more errors than a
bounded subspace distance decoder.

Given the layered structure of $\mathcal{E}$, our decoding algorithm for $\mathcal{E}$ is based on a decoding
algorithm for $\mathcal{E}^k$, shown below in Algorithm~\ref{alg:Ek}, for any $k$. We denote the
codewords in $\mathcal{E}^0$ as $E_{0,j}^0$ for $0 \leq j \leq
|\mathcal{E}^0|-1$.


\begin{algorithm}\label{alg:Ek}
\renewcommand{\labelenumi}{\ref{alg:Ek}.\theenumi}
$\mathrm{EBDD}(k, {\bf A})$.\\
Input: $k$ and ${\bf A} = ({\bf A}_1 | {\bf A}_2 | {\bf A}_3) \in
\mathrm{GF}(q)^{a \times n}$, ${\bf A}_1 \in \mathrm{GF}(q)^{a
\times (r - kd)}$,
${\bf A}_2 \in \mathrm{GF}(q)^{a \times kd}$, ${\bf A}_3 \in \mathrm{GF}(q)^{a \times (n-r)}$.\\
Output: $(E_{i,j}^k, d_k, f_k)$.
\begin{enumerate}
\item \label{step:E0} If $k=0$, use the decoder for $\mathcal{E}^0$
to obtain $E_{0,j}^0$, calculate $d_k = \ds(R({\bf A}), E_{0,j}^0)$,
and return $(E_{0,j}^0, d_k, 0)$. If the decoder returns a failure,
return $(I({\bf 0}), d, 0)$.

\item \label{step:decode_Ck} Use the decoder of $I(\mathcal{C}^k)$ on $({\bf A}_1 | {\bf A}_2)$ to obtain ${\bf C}_i^k$.
If the decoder returns a failure, set ${\bf C}_i^k = {\bf 0}$, ${\bf
D}_j^k = {\bf 0}$ and return $(E_{i,j}^k,
d, 0)$.

\item \label{step:decode_Dk} Use the decoder of $I(\mathcal{D}^k)$
on $({\bf A}_1 | {\bf A}_3)$ to obtain ${\bf D}_j^k$. If the decoder
returns a failure, set ${\bf D}_j^k = {\bf 0}$ and return $(E_{i,j}^k, d, 0)$.

\item \label{step:d0} Calculate $d_k = \ds(R({\bf A}), E_{i,j}^k)$ and
$f_k = 2d- \max\{\ds(R({\bf A}_1 | {\bf A}_2), I({\bf C}_i^k)),
\ds(R({\bf A}_1 | {\bf A}_3), I({\bf D}_j^k))\}$ and return
$(E_{i,j}^k, d_k, f_k)$.
\end{enumerate}
\end{algorithm}

Algorithm~\ref{alg:Ek} is based on the bounded distance decoder
proposed in \cite{koetter_it08}. When $k=0$, $\mathcal{E}^0$ is
simply a KK code, and the algorithm in \cite{koetter_it08} is used
directly; when $k \geq 1$, given the structure of $\mathcal{E}^k$,
two decoding attempts are made based on $({\bf A}_1 | {\bf A}_2)$
and $({\bf A}_1 | {\bf A}_3)$, and both are based on the decoding
algorithm in \cite{koetter_it08}.

We remark that Algorithm~\ref{alg:Ek} always return $(E_{i,j}^k,
d_k, f_k)$. If a unique nearest codeword in $\mathcal{E}^k$ at
distance no more than $d-1$ from $R({\bf A})$ exists, then by
Lemma~\ref{lemma:truncate} Steps~\ref{alg:Ek}.\ref{step:decode_Ck}
and \ref{alg:Ek}.\ref{step:decode_Dk} succeed and
Algorithm~\ref{alg:Ek} returns the unique nearest codeword in
$E_{i,j}^k$. However, when such unique codeword in $\mathcal{E}^k$
at distance no more than $d-1$ does not exist, the return value
$f_k$ can be used to find the unique nearest codeword because $f_k$
is a lower bound on the distance from the received subspace to any
other codeword in $\mathcal{E}^k$. Also, when $f_k=0$,
Algorithm~\ref{alg:E} below always returns a failure. Thus, we call
Algorithm~\ref{alg:Ek} an enhanced bounded distance decoder.

\begin{lemma} \label{lemma:fk}
Suppose the output of $\mathrm{EBDD}(k, {\bf A})$ is $(E_{i,j}^k,
d_k, f_k)$, then $\ds(R({\bf A}), E_{u,v}^k) \geq f_k$ for any $E_{u,v}^k \in \mathcal{E}^k$ provided $(u,v) \neq
(i,j)$.
\end{lemma}

\begin{proof}
The case $f_k = 0$ is trivial, and it suffices to consider $f_k =
\min\{2d- \ds(R({\bf A}_1 | {\bf A}_2), I({\bf C}_i^k)), 2d-
\ds(R({\bf A}_1 | {\bf A}_3), I({\bf D}_j^k))\}$. When $u \neq i$,
Lemma \ref{lemma:truncate} yields
\begin{eqnarray}
    \nonumber
    \ds(R({\bf A}), E_{u,v}^k) &\geq& \ds(R({\bf A}_1|{\bf A}_2), I({\bf
    C}_u^k))\\
    \nonumber
    &\geq& \ds(I({\bf C}_i^k), I({\bf C}_u^k)) - \ds(R({\bf A}_1|{\bf A}_2), I({\bf
    C}_i^k))\\
    \nonumber
    &\geq& 2d- \ds(R({\bf A}_1|{\bf A}_2), I({\bf C}_i^k)) \geq f_k.
\end{eqnarray}
Similarly, when $v \neq j$, we obtain $\ds(R({\bf A}), E_{u,v}^k)
\geq 2d- \ds(R({\bf A}_1 | {\bf A}_3), I({\bf D}_j^k)) \geq f_k$.
\end{proof}

The algorithm for $\mathcal{E}$ thus follows.

\begin{algorithm}\label{alg:E}
\renewcommand{\labelenumi}{\ref{alg:E}.\theenumi}
Decoder for $\mathcal{E}$.\\
Input: ${\bf A} = ({\bf A}_0 | {\bf A}_3) \in \mathrm{GF}(q)^{a \times n}$,
${\bf A}_0 \in \mathrm{GF}(q)^{a \times r}$, ${\bf A}_3 \in \mathrm{GF}(q)^{a \times (n-r)}$.\\
Output: Either a failure or the unique nearest codeword in
$\mathcal{E}$ from $R({\bf A})$.
\begin{enumerate}
\item \label{step:t} If $\rk({\bf A}) < r-d+1$, return a failure.
\item \label{step:t} Calculate $r - \rk({\bf A}_0) = ld + m$ where $0 \leq l \leq \left\lfloor \frac{r}{d} \right\rfloor$ and $0 \leq m < d$.

\item \label{step:Ek} Call $\mathrm{EBDD}(l, {\bf A})$ to obtain $(E_{i,j}^l,d_l,f_l)$.
If $d_l \leq d-1$, return $E_{i,j}^l$.

\item \label{step:Ek+1} If $m=0$, return a failure. Otherwise, call $\mathrm{EBDD}(l+1, {\bf A})$ to obtain $(E_{s,t}^{l+1}, d_{l+1}, f_{l+1})$.
If $d_{l+1} \leq d-1$, return $E_{s,t}^{l+1}$.

\item \label{step:compare_d0} If $d_l < \min\{d+m, f_l, d_{l+1}, f_{l+1}, 2d-m\}$,
return $E_{i,j}^l$. If $d_{l+1} < \min\{d+m, d_l, f_l, f_{l+1},
2d-m\}$, return $E_{s,t}^{l+1}$.
\item \label{step:return} Return a failure.
\end{enumerate}
\end{algorithm}

\begin{proposition} \label{prop:decoding_E}
If the received subspace is at subspace distance at most $d-1$ from
a codeword in $\mathcal{E}$, then Algorithm~\ref{alg:E} returns this
codeword. Otherwise, Algorithm~\ref{alg:E} returns either a failure
or the unique codeword closest to the received subspace in the subspace metric.
\end{proposition}

\begin{proof}
We first show that Algorithm~\ref{alg:E} returns the unique nearest
codeword in $\mathcal{E}$ to the received subspace if it is at
subspace distance at most $d-1$. For all $1 \leq k \leq \left\lfloor
\frac{r}{d} \right\rfloor$ and $E_{u,v}^k \in \mathcal{E}^k$, Lemma
\ref{lemma:truncate} and (\ref{eq:ds_bound}) yield
\begin{equation}\label{eq:d(R(A),El)}
    \ds(R({\bf A}), E_{u,v}^k) \geq \ds(R({\bf A}_0), I({\bf
    C}_u^k)) \geq |r-kd - \rk({\bf A}_0)| = |(l-k)d + m|.
\end{equation}
Similarly (\ref{eq:ds_bound}) yields $\ds(R({\bf A}), E_{0,v}^0)
\geq ld+m$ for any $v$. Hence $\ds(R({\bf A}), \mathcal{E}^k) \geq
d$ for $k \leq l-1$ or $k \geq l+2$. Therefore, the unique nearest
codeword is either in $\mathcal{E}^l$ or $\mathcal{E}^{l+1}$ and
applying Algorithm~\ref{alg:Ek} for $\mathcal{E}^l$ and
$\mathcal{E}^{l+1}$ always returns the nearest codeword.

We now show that when the distance from the received subspace to the
code is at least $d$, Algorithm~\ref{alg:E} either produces the
unique nearest codeword or returns a failure. First, by
(\ref{eq:d(R(A),El)}), $\ds(R({\bf A}), \mathcal{E}^{l-1}) = d+m$
and $\ds(R({\bf A}), \mathcal{E}^{l+2}) = 2d-m$, while $\ds(R({\bf
A}), \mathcal{E}^k) \geq 2d$ for $k \leq l-2$ or $k \geq l+3$. Also,
by Lemma \ref{lemma:fk}, $\ds(R({\bf A}), E_{u,v}^l) \geq f_l$ for
all $(u,v) \neq (i,j)$ and $\ds(R({\bf A}), \mathcal{E}^{l+1}) \geq
\min\{d_{l+1}, f_{l+1}\}$. Therefore, if $d_l < \min\{d+m, f_l,
d_{l+1}, f_{l+1}, 2d-m\}$, then $E_{i,j}^l$ is the unique
codeword  closest to $R({\bf A})$. Similarly, if $d_{l+1} < \min\{d+m, d_l,
f_l, f_{l+1}, 2d-m\}$, then $E_{s,t}^{l+1}$ is the unique
codeword  closest to $R({\bf A})$.
\end{proof}

We note that when $\rk({\bf A}) < r-d+1$, by (\ref{eq:ds_bound})
Steps~\ref{alg:Ek}.\ref{step:decode_Ck} and
\ref{alg:Ek}.\ref{step:decode_Dk} would both fail, and
Algorithm~\ref{alg:E} will return a failure. We also justify why
Algorithm~\ref{alg:E} returns a failure if $d_l \geq d$ and $m=0$ in
Step \ref{alg:E}.\ref{step:Ek}. Suppose $d_l \geq d$ and $m = 0$ and
we apply Algorithm~\ref{alg:Ek} for $\mathcal{E}^{l+1}$. Then we
have $d_l \geq d+m$ and by (\ref{eq:d(R(A),El)}) $d_{l+1} \geq |d-m|
= d+m$. Therefore, neither inequality in Step
\ref{alg:E}.\ref{step:compare_d0} is satisfied and the decoder
returns a failure.

By Proposition~\ref{prop:decoding_E}, Algorithm~\ref{alg:E} decodes
beyond the half distance. However, the decoding radius of
Algorithm~\ref{alg:E} is limited. It is easy to see that the
decoding radius of Algorithm~\ref{alg:E} is at most $d+\left \lfloor
\frac{d}{2}\right\rfloor$ due to the terms $d+m$ and $2d-m$ in the
inequalities in Step~\ref{alg:E}.\ref{step:compare_d0}. We emphasize
that this is just an upper bound, and its tightness is unknown.
Suppose $r - \rk({\bf A}_0) = ld + m$, when Algorithm~\ref{alg:E}
decodes beyond half distance, it is necessary that $f_l$ and
$f_{l+1}$ be both nonzero in Step~\ref{alg:E}.\ref{step:compare_d0}.
This implies that the row space of $({\bf A}_1 | {\bf A}_2)$ is at
subspace distance no more than $d-1$ from $I(\mathcal{C}^l)$ and
$I(\mathcal{C}^{l+1})$ and that the row spaces of $({\bf A}_1 | {\bf
A}_3)$ are at subspace distance no more than $d-1$ from
$I(\mathcal{D}^l)$ and $I(\mathcal{D}^{l+1})$.

We note that the inequalities in
Step~\ref{alg:E}.\ref{step:compare_d0} are strict in order to ensure
that the output of the decoder is the \textbf{unique} nearest
codeword from the received subspace. However, if one of the nearest
codewords is an acceptable outcome, then equality can be included in
the inequalities in Step~\ref{alg:E}.\ref{step:compare_d0}.

Our decoding algorithm can be readily simplified in order to obtain
a bounded subspace distance decoder, by removing Step \ref{alg:E}.\ref{step:compare_d0}.
We
emphasize that the general decoding algorithm has the same order of
complexity as this simplified bounded subspace distance decoding
algorithm.

Finally, we note that the decoding algorithms and discussions above
consider the subspace metric. It is also remarkable that our decoder
remains the same if the injection metric is used instead. We
formalize this by the following proposition.

\begin{proposition}\label{prop:decoding_I}
If the received subspace is at injection distance at most $d-1$ from
a codeword in $\mathcal{E}$, then Algorithm~\ref{alg:E} returns this
codeword. Otherwise, Algorithm~\ref{alg:E} returns either a failure
or the unique codeword closest to the received subspace in the injection metric.
\end{proposition}

The proof of Proposition~\ref{prop:decoding_I} is based on the
observation that a codeword in a CDC is closest to the received
subspace in the subspace metric if and only if the codeword is
closest to the received subspace in the injection metric by
(\ref{eq:di}), and is hence omitted.

The complexity of the bounded subspace distance decoder in
\cite{koetter_it08} for a KK code in $E(q,n)$ is on the order of
$O(n^2)$ operations over $\mathrm{GF}(q)^{n-r}$ for $r \leq
\frac{n}{2}$, which is hence the complexity of decoding
$\mathcal{E}^0$. This algorithm can be easily generalized to include
the case where $r
> \frac{n}{2}$, and we obtain a complexity on the order of $O(n^2)$
operations over $\mathrm{GF}(q^{\max\{r,n-r\}})$. Thus the
complexity of decoding $I(\mathcal{C}^k)$ and $I(\mathcal{D}^k)$ for
$k \geq 1$ is on the order of $O(r^2)$ operations over
$\mathrm{GF}(q^{\max\{kd,r-kd\}})$ and $O((n-kd)^2)$ operations over
$\mathrm{GF}(q^{\max\{r,n-kd-r\}})$, respectively. The complexity of
the decoding algorithm for $\mathcal{E}^k$ is on the order of the
maximum of these two quantities. It is easily shown that the
complexity is maximized for $k=0$, that is, our decoding algorithm
has the same order of complexity as the algorithm for the KK code
$\mathcal{E}^0$.

\section{Covering properties of CDCs}\label{sec:covering}
The packing properties of CDCs have been studied in
\cite{koetter_it08, xia_dcc09, skachek_arxiv08, gabidulin_isit08,
kohnert_mmics08} and an asymptotic packing rate of CDCs was defined
and determined in \cite{koetter_it08}. Henceforth in this section,
we focus on the covering properties of CDCs in the Grassmannian instead. We emphasize
that since $\ds(U,V) = 2\di(U,V)$ for all $U,V \in E_r(q,n)$, we
consider only the injection distance in this section. Furthermore,
since $\di(U,V)=\di(U^\perp,V^\perp)$ for all $U,V \in E_r(q,n)$,
without loss of generality we assume that $r\leq \left\lfloor
\frac{n}{2} \right\rfloor$ in this section.

\subsection{Properties of balls in the
Grassmannian}\label{sec:balls_CDC}

We first investigate the properties of balls in the Grassmannian
$E_r(q,n)$, which will be instrumental in our study of covering
properties of CDCs. First, we derive bounds on the volume of balls
in $E_r(q,n)$.

\begin{lemma}\label{lemma:bounds_Vc}
For all $q$, $n$, $r \leq \left\lfloor \frac{n}{2} \right\rfloor$,
and $0 \leq t \leq r$, $q^{t(n-t)} \leq \Vc(t) < K_q^{-2}
q^{t(n-t)}$.
\end{lemma}

\begin{proof}
First, we have $\Vc(t) \geq \Nc(t) \geq q^{t(n-t)}$ by
(\ref{eq:Gaussian}). Also, $\Nc(d) < K_q^{-1} \Nr(q,n-r,r,d)$, and
hence $\Vc(t) < K_q^{-1} \Vr(q,n-r,r,t) < K_q^{-2} q^{t(n-t)}$ as
$\Vr(q,n-r,r,t) < K_q^{-1} q^{t(n-t)}$ \cite[Lemma
9]{gadouleau_it08_dep}.
\end{proof}

We now determine the volume of the intersection of two
\textbf{spheres} of radii $u$ and $s$ respectively and distance $d$
between their centers, which is referred to as the intersection
number $\Jc(u,s,d)$ of the association scheme \cite{brouwer_book89}.
The intersection number is an important parameter of an association
scheme.

\begin{lemma}\label{lemma:Jc}
For all $u$, $s$, and $d$ between $0$ and $r$,
\begin{equation}
    \nonumber
    \Jc(u,s,d) = \frac{1}{{n \brack r} \Nc(d)}
    \sum_{i=0}^r \mu_i E_u(i) E_s(i) E_d(i),
\end{equation}
where $\mu_i = {n \brack i} - {n \brack i-1}$ and $E_j(i)$ is a
$q$-Eberlein polynomial \cite{delsarte_siam76}:
\begin{equation}
    \nonumber
    E_j(i) = \sum_{l=0}^j (-1)^{j-l} q^{li + {j-l \choose 2}} {r-l \brack
    r-j} {r-l \brack i} {n-r+l-i \brack l}.
\end{equation}
\end{lemma}


Although Lemma~\ref{lemma:Jc} is obtained by a direct application of
Theorems 3.5 and 3.6 in \cite[Chapter II]{bannai_book83}, we present
it formally here since it is a fundamental geometric property of the
Grassmannian and is very instrumental in our study of CDCs. We also
obtain a recursion formula for $\Jc(u,s,d)$.

\begin{lemma}\label{lemma:recursion_Jc}
$\Jc(u,s,d)$ satisfies the following recursion: $\Jc(0,s,d) =
\delta_{s,d}$, $\Jc(u,0,d) = \delta_{u,d}$, and
\begin{equation}
    \nonumber
    c_{u+1} \Jc(u+1,s,d) = b_{s-1} \Jc(u,s-1,d) + (a_s - a_u) \Jc(u,s,d)
    + c_{s+1} \Jc(u,s+1,d) - b_{u-1} \Jc(u-1,s,d),
\end{equation}
where $c_j = \Jc(1,j-1,j) = {j \brack 1}^2$, $b_j = \Jc(1,j+1,j) =
q^{2j+1} {r-j \brack 1} {n-r-j \brack 1}$, and $a_j = \Jc(1,j,j) =
\Nc(1) - b_j - c_j$ for $0 \leq j \leq r$.
\end{lemma}

The proof follows directly from \cite[Lemma 4.1.7]{brouwer_book89},
\cite[Theorem 9.3.3]{brouwer_book89}, and  \cite[Chapter 4,
(1a)]{brouwer_book89}, and hence is omitted. Let $\Ic(u,s,d)$ denote
the intersection of two \textbf{balls} in $E_r(q,n)$ with radii $u$
and $s$ and distance $d$ between their centers. Since $\Ic(u,s,d) =
\sum_{i=0}^u \sum_{j=0}^s \Jc(i,j,d)$, Lemma~\ref{lemma:Jc} also
leads to an analytical expression for $\Ic(u,s,d)$.
Proposition~\ref{prop:inter_2_balls} below shows that $\Ic(u,s,d)$
decreases as $d$ increases.

\begin{proposition}\label{prop:inter_2_balls}
For all $u$ and $s$, $\Ic(u,s,d)$ is a non-increasing function of
$d$.
\end{proposition}

The proof of Proposition~\ref{prop:inter_2_balls} is given in
Appendix~\ref{app:prop:inter_2_balls}. Therefore, the minimum
nonzero intersection between two balls with radii $u$ and $s$ in
$E_r(q,n)$ is given by $\Ic(u,s,u+s) = \Jc(u,s,u+s)$ for $u+s \leq
r$. By Lemma~\ref{lemma:recursion_Jc}, it is easily shown that
$\Jc(u,s,u+s) = {u+s \brack u}^2$ for all $u$ and $s$ when $u+s \leq
r$.

We derive below an upper bound on the union of balls in $E_r(q,n)$
with the same radius.

\begin{lemma}\label{lemma:B}
The volume of the union of \emph{any} $K$ balls in $E_r(q,n)$ with
radius $\rho$ is at most
\begin{eqnarray}
    \nonumber
    \Bc(K, \rho) &&=  K\Vc(\rho)   - \sum_{a=1}^l [\Ac(q,n,r,r-a+1) -
    \Ac(q,n,r,r-a+2)]  \Ic(\rho,\rho,r-a+1)\\
    \label{eq:B}
    && - [K - \Ac(q,n,r,r-l+1)]  \Ic(\rho,\rho,r-l),
\end{eqnarray}
where $l = \max\{a: K \geq \Ac(q,n,r,r-a+1)\}$.
\end{lemma}

\begin{proof}
Let $\{ U_i \}_{i=0}^{K-1}$ denote the centers of $K$ balls with
radius $\rho$ and let $\mathcal{V}_j = \{ U_i \}_{i=0}^{j-1}$ for $1
\leq j \leq K$. Without loss of generality, we assume that the
centers are labeled such that $\di(U_j, \mathcal{V}_j)$ is
non-increasing for $j \geq 1$. For $1 \leq a \leq l$ and
$\Ac(q,n,r,r-a+2) \leq j < \Ac(q,n,r,r-a+1)$, we have $\di(U_j,
\mathcal{V}_j) = \di(\mathcal{V}_{j+1}) \leq r-a+1$. By
Proposition~\ref{prop:inter_2_balls}, $U_j$ hence covers at most
$\Vc(\rho) - \Ic(\rho, \rho, r-a+1)$ subspaces that are not
previously covered by balls centered at $\mathcal{V}_j$.
\end{proof}

We remark that using any upper bound on $\Ac(q,n,r,r-a+1)$ in the
proof of Lemma~\ref{lemma:B} leads to a valid upper bound on $\Bc(K,
\rho)$. Hence, although the value of $\Ac(q,n,r,r-a+1)$ is unknown
in general, the upper bound in (\ref{eq:bounds_Ac}) can be used in
(\ref{eq:B}) in order to obtain an upper bound on the volume of the
union on balls in the Grassmannian.

\subsection{Covering CDCs}\label{sec:covering_CDC}
The {\em covering radius} of a CDC $\mathcal{C} \subseteq E_r(q,n)$
is defined as $\rho = \max_{U \in E_r(q,n)} \di(U,\mathcal{C})$. We
denote the minimum cardinality of a CDC in $E_r(q,n)$ with covering
radius $\rho$ as $\Kc(q,n,r,\rho)$. Since $\Kc(q,n,n-r,\rho) =
\Kc(q,n,r,\rho)$, we assume $r \leq \left\lfloor \frac{n}{2}
\right\rfloor$. Also, $\Kc(q,n,r,0) = {n \brack r}$ and
$\Kc(q,n,r,r) = 1$, hence we assume $0 < \rho < r$ henceforth. We
first derive lower bounds on $\Kc(q,n,r,\rho)$.

\begin{lemma}\label{lemma:bound_B}
For all $q$, $n$, $r \leq \left\lfloor \frac{n}{2} \right\rfloor$,
and $0 < \rho < r$, $\Kc(q,n,r,\rho) \geq \min \left\{K : \Bc(K,
\rho) \geq {n \brack r} \right\} \geq \frac{{n \brack
r}}{\Vc(\rho)}$.
\end{lemma}

\begin{proof}
Let $\mathcal{C}$ be a CDC with cardinality $\Kc(q,n,r,\rho)$ and
covering radius $\rho$. Then the balls around the codewords cover
the ${n \brack r}$ subspaces in $E_r(q,n)$; however, by
Lemma~\ref{lemma:B}, they cannot cover more than $\Bc(|\mathcal{C}|,
\rho)$ subspaces. Therefore, $\Bc(\Kc(q,n,r,\rho), \rho) \geq {n
\brack r}$ and we obtain the first inequality. Since $\Bc(K, \rho)
\leq K \Vc(\rho)$ for all $K$, we obtain the second inequality.
\end{proof}

The second lower bound in Lemma \ref{lemma:bound_B} is referred to
as the sphere covering bound for CDCs. This bound can also be
refined by considering the distance distribution of a covering code.

\begin{proposition}\label{prop:linear_inequalities}
For $0 \leq \delta \leq \rho$, let $T_\delta = \min \sum_{i=0}^r
A_i(\delta)$, where the minimum is taken over all integer sequences
$\{A_i(\delta)\}$ which satisfy $A_i(\delta) = 0$ for $0 \leq i \leq
\delta-1$, $1 \leq A_\delta(\delta) \leq \Nc(\delta)$, $0 \leq
A_i(\delta) \leq \Nc(i)$ for $\delta+1 \leq i \leq r$, and
$\sum_{i=0}^r A_i(\delta) \sum_{s=0}^\rho \Jc(l,s,i) \geq \Nc(l)$
for $0 \leq l \leq r$.
Then $\Kc(q,n,r,\rho) \geq \max_{0 \leq \delta \leq \rho} T_\delta$.
\end{proposition}

\begin{proof}
Let $\mathcal{C}$ be a CDC with covering radius $\rho$. For any $U
\in E_r(q,n)$ at distance $\delta$ from $\mathcal{C}$, let
$A_i(\delta)$ denote the number of codewords at distance $i$ from
$U$. Then $\sum_{i=0}^r A_i(\delta) = |\mathcal{C}|$ and we easily
obtain $A_i(\delta) = 0$ for $0 \leq i \leq \delta-1$, $1 \leq
A_\delta(\delta) \leq \Nc(\delta)$, and $0 \leq A_i(\delta) \leq
\Nc(i)$ for $\delta+1 \leq i \leq r$. Also, for $0 \leq l \leq r$,
all the subspaces at distance $l$ from $U$ are covered, hence
$\sum_{i=0}^r A_i(\delta) \sum_{s=0}^\rho \Jc(l,s,i) \geq \Nc(l)$.
\end{proof}

We remark that Proposition~\ref{prop:linear_inequalities} is a
tighter lower bound than the sphere covering bound. However,
determining $T_\delta$ is computationally infeasible for large
parameter values.

Another set of linear inequalities is obtained from the inner
distribution $\{a_i\}$ of a covering code $\mathcal{C}$, defined as
$a_i \df \frac{1}{|\mathcal{C}|} \sum_{C \in \mathcal{C}} |\{D \in
\mathcal{C} : \di(C,D) = i\}|$ for $0 \leq i \leq r$
\cite{delsarte_it98}.

\begin{proposition}\label{prop:linear_ai}
Let $t = \min \sum_{i=0}^r a_i$, where the minimum is taken over all
sequences $\{a_i\}$ satisfying $a_0 = 1$, $0 \leq a_i \leq \Nc(i)$
for $1 \leq i \leq r$, $\sum_{i=0}^r a_i \sum_{s=0}^\rho \Jc(l,s,i)
\geq \Nc(l)$ for $0 \leq l \leq r$, and $\sum_{i=0}^r a_i
\frac{E_i(l)}{\Nc(i)} \geq 0$ for $0 \leq l \leq r$.
Then $\Kc(q,n,r,\rho) \geq t$.
\end{proposition}

\begin{proof}
Let $\mathcal{C}$ be a CDC with covering radius $\rho$ and inner
distribution $\{a_i\}$. Proposition~\ref{prop:linear_inequalities}
yields $0 \leq a_i \leq \Nc(i)$ for $1 \leq i \leq r$, $\sum_{i=0}^r
a_i \sum_{s=0}^\rho \Jc(l,s,i) \geq \Nc(l)$ for $0 \leq l \leq r$,
while $a_0 = 1$ follows the definition of $a_i$. By the generalized
MacWilliams inequalities \cite[Theorem 3]{delsarte_it98},
$\sum_{i=0}^r a_i F_l(i) \geq 0$, where $F_l(i) =
\frac{\mu_l}{\Nc(i)} E_i(l)$ are the q-numbers of the association
scheme \cite[(15)]{delsarte_it98}, which yields $\sum_{i=0}^r a_i
\frac{E_i(l)}{\Nc(i)} \geq 0$. Since $\sum_{i=0}^r a_i =
|\mathcal{C}|$ we obtain that $|\mathcal{C}| \geq t$.
\end{proof}

Lower bounds on covering codes with the Hamming metric can be
obtained through the concept of the excess of a code
\cite{vanwee_jct91}. This concept being independent of the
underlying metric, it was adapted to the rank metric in
\cite{gadouleau_it08_covering}. We adapt it to the injection metric
for CDCs below, thus obtaining the lower bound in
Proposition~\ref{prop:excess_bound}.

\begin{proposition}\label{prop:excess_bound}
For all $q$, $n$, $r \leq \left\lfloor \frac{n}{2} \right\rfloor$,
and $0 < \rho < r$, $ \Kc(q,n,r,\rho) \geq \frac{{n \brack r}}
{\Vc(\rho) - \frac{\epsilon}{\delta}\Nc(\rho)}$, where $\epsilon \df
\left\lceil \frac{b_\rho}{c_{\rho+1}} \right\rceil c_{\rho+1}
-b_\rho$, $\delta \df \Nc(1) - c_\rho + 2\epsilon$, and $b_\rho$ and
$c_{\rho+1}$ are defined in Lemma \ref{lemma:recursion_Jc}.
\end{proposition}

The proof of Proposition \ref{prop:excess_bound} is given in
Appendix \ref{app:prop:excess_bound}. We now derive upper bounds on
$\Kc(q,n,r,\rho)$. First, we investigate how to expand covering
CDCs.

\begin{lemma}\label{lemma:Kc(rho+1)}
For all $q$, $n$, $r \leq \left\lfloor \frac{n}{2} \right\rfloor$,
and $0 < \rho < r$, $\Kc(q,n,r,\rho) \leq \Kc(q,n-1,r,\rho-1) \leq
{n-\rho \brack r}$, and $\Kc(q,n,r,\rho) \leq \Kc(q,n,r-1,\rho-1)
\leq {n \brack r-\rho}$.
\end{lemma}

The proof of Lemma \ref{lemma:Kc(rho+1)} is given in Appendix
\ref{app:lemma:Kc(rho+1)}. The next upper bound is a straightforward
adaptation of \cite[Proposition 12]{gadouleau_it08_covering}.

\begin{proposition}\label{prop:bound_combinatorial}
For all $q$, $n$, $r \leq \left\lfloor \frac{n}{2} \right\rfloor$,
and $0 < \rho < r$, $\Kc(q,n,r,\rho) \leq   \left\{ 1-\log_{n \brack
r} \left({n \brack r} - \Vc(\rho) \right) \right\}^{-1} + 1$.
\end{proposition}

The proof of Proposition \ref{prop:bound_combinatorial} is given in
Appendix \ref{app:prop:bound_combinatorial}. The next bound is a
direct application of \cite[Theorem 12.2.1]{cohen_book97}.

\begin{proposition}\label{prop:bound_JSL}
For all $q$, $n$, $r \leq \left\lfloor \frac{n}{2} \right\rfloor$,
and $0 < \rho < r$, $\Kc(q,n,r,\rho) \leq \frac{{n \brack
r}}{\Vc(\rho)} \left\{ 1 + \ln \Vc(\rho) \right\}$.
\end{proposition}

The bound in Proposition \ref{prop:bound_JSL} can be refined by
applying the greedy algorithm described in \cite{clark_ejc97} to
CDCs.


\begin{proposition}\label{prop:bound_domination}
Let $k_0$ be the cardinality of an augmented KK code with minimum
distance $2\rho+1$ in $E_r(q,n)$ for $2 \rho < r$ and $k_0 = 1$ for
$2 \rho \geq r$. Then for all $k \geq k_0$, there exists a CDC with
cardinality $k$ which covers at least ${n \brack r} - u_k$
subspaces, where $u_{k_0} \df {n \brack r} - k_0 \Vc(\rho)$ and
$u_{k+1} = u_k - \left\lceil \frac{u_k \Vc(\rho)}
    {\min \left\{ {n \brack r} - k, \Bc(u_k, \rho) \right\}}
    \right\rceil$ for all $k \geq k_0$. Thus $\Kc(q,n,r,\rho) \leq \min\{k : u_k =
0\}$.
\end{proposition}

The proof of Proposition~\ref{prop:bound_domination} is given in
Appendix~\ref{app:prop:bound_domination}.

Using the bounds derived above, we finally determine the asymptotic
behavior of $\Kc(q,n,r,\rho)$. The rate of a covering CDC
$\mathcal{C} \subseteq E_r(q,n)$ is defined as $\frac{\log_q
|\mathcal{C}|}{ \log_q |E_r(q,n)|}$.
We remark that this rate is defined in a combinatorial sense: the rate describes
how well a CDC covers the Grassmannian. We use the following
normalized parameters: $r' = \frac{r}{n}$, $\rho' = \frac{\rho}{n}$,
and the asymptotic rate $\kc(r',\rho') = \lim \inf_{n \rightarrow
\infty} \frac{\log_q \Kc(q,n,r,\rho)}{\log_q {n \brack r}}$.

\begin{proposition}\label{prop:kc}
For all $0 \leq \rho' \leq r' \leq \frac{1}{2}$, $\kc(r',\rho') = 1
- \frac{\rho'(1-\rho')}{r'(1-r')}$.
\end{proposition}

\begin{proof}
The bounds on $\Vc(\rho)$ in Lemma \ref{lemma:bounds_Vc} together
with the sphere covering bound yield $\Kc(q,n,r,\rho) > K_q^2
q^{r(n-r) - \rho(n-\rho)}$. Using the bounds on the Gaussian
polynomial in Section~\ref{sec:CDCs_and_rank_metric}, we obtain
$\kc(r',\rho') \geq 1 - \frac{\rho'(1-\rho')}{r'(1-r')}$. Also,
Proposition \ref{prop:bound_JSL} leads to $\Kc(q,n,r,\rho) <
K_q^{-1} q^{r(n-r) - \rho(n-\rho)} [1 + \ln (K_q^{-2}) +
\rho(n-\rho)\ln q]$, which asymptotically becomes $\kc(r',\rho')
\leq 1 - \frac{\rho'(1-\rho')}{r'(1-r')}$.
\end{proof}

The proof of Proposition \ref{prop:kc} indicates that
$\Kc(q,n,r,\rho)$ is on the order of $q^{r(n-r) - \rho(n-\rho)}$.

We finish this section by studying the covering properties of
liftings of rank metric codes. We first prove that they have maximum
covering radius.

\begin{lemma} \label{lemma:covering_lifting}
Let $I(\mathcal{C}) \subseteq E_r(q,n)$ be the lifting of a rank
metric code in $\mathrm{GF}(q)^{r \times (n-r)}$. Then
$I(\mathcal{C})$ has covering radius $r$.
\end{lemma}

\begin{proof}
Let $D \in E_r(q,n)$ be generated by $({\bf 0} | {\bf D}_1)$, where
${\bf D}_1 \in \mathrm{GF}(q)^{r \times (n-r)}$ has rank $r$. Then,
for any codeword $I({\bf C})$ generated by $({\bf I}_r | {\bf C})$,
it is easily seen that $\di(D, I({\bf C})) = \di(R({\bf 0}), R({\bf I}_r)) = r$ by Lemma~\ref{lemma:truncate}.
\end{proof}

Lemma~\ref{lemma:covering_lifting} is significant for the design of
CDCs. It is shown in \cite{koetter_it08} that liftings of rank
metric codes can be used to construct nearly optimal packing CDCs.
However, Lemma~\ref{lemma:covering_lifting} indicates that for any
lifting of a rank metric code, there exists a subspace at distance
$r$ from the code. Hence, adding this subspace to the code leads to
a supercode with higher cardinality and the same minimum distance
since $d \leq r$. Thus an optimal CDC cannot be designed from a
lifting of a rank metric code.

Although liftings of rank metric codes have poor covering
properties, below we construct a class of covering CDCs by using
permuted liftings of rank metric covering codes. We thus relate the
minimum cardinality of a covering CDC to that of a covering code
with the rank metric. For all $n$ and $r$, we denote the set of
subsets of $\{0,1,\ldots,n-1\}$ with cardinality $r$ as $S_n^r$. For
all $J \in S_n^r$ and all ${\bf C} \in \mathrm{GF}(q)^{r \times
(n-r)}$, let $I(J,{\bf C}) = R(\pi({\bf I}_r | {\bf C})) \in
E_r(q,n)$, where $\pi$ is the permutation of $\{0, 1, \ldots, n-1\}$
satisfying $J = \{ \pi(0),  \pi(1), \ldots, \pi(r-1)\}$, $\pi(0) <
\pi(1) < \ldots < \pi(r-1)$, and $\pi(r) < \pi(r+1) < \ldots <
\pi(n-1)$. We remark that $\pi$ is uniquely determined by $J$. It is
easily shown that $\di(I(J,{\bf C}), I(J,{\bf D})) = \dr({\bf C},
{\bf D})$ for all $J \in S_n^r$ and all ${\bf C}, {\bf D} \in
\mathrm{GF}(q)^{r \times (n-r)}$.

\begin{proposition}\label{prop:Ks<Kr}
For all $q$, $n$, $r \leq \left\lfloor \frac{n}{2} \right\rfloor$,
and $0 < \rho < r$, $\Kc(q,n,r,\rho) \leq {n \choose r}
\Kr(q^{n-r},r,\rho)$.
\end{proposition}

\begin{proof}
Let $\mathcal{C} \subseteq \mathrm{GF}(q)^{r \times (n-r)}$ have
rank covering radius $\rho$ and cardinality $\Kr(q^{n-r},r,\rho)$.
We show below that $L(\mathcal{C}) = \{I(J,{\bf C}) : J \in S_n^r,
{\bf C} \in \mathcal{C}\}$ is a CDC with covering radius $\rho$. Any
$U \in E_r(q,n)$ can be expressed as $I(J,{\bf V})$ for some $J \in
S_n^r$ and some ${\bf V} \in \mathrm{GF}(q)^{r \times (n-r)}$. Also,
by definition, there exists ${\bf C} \in \mathcal{C}$ such that
$\dr({\bf C}, {\bf V}) \leq \rho$ and hence $\di(U,I(J,{\bf C})) =
\dr({\bf C}, {\bf V}) \leq \rho$. Thus $L(\mathcal{C})$ has covering
radius $\rho$ and cardinality $\leq {n \choose r}
\Kr(q^{n-r},r,\rho)$.
\end{proof}

It is shown in \cite{gadouleau_it08_covering} that for $r \leq n-r$,
$\Kr(q^{n-r}, r, \rho)$ is on the order of $q^{r(n-r) -
\rho(n-\rho)}$, which is also the order of $\Kc(q,n,r,\rho)$. The bound in
Proposition~\ref{prop:Ks<Kr} is relatively tighter for large $q$
since ${n \choose r}$ is independent of $q$.


\appendix

\subsection{Proof of
Proposition~\ref{prop:inter_2_balls}}\label{app:prop:inter_2_balls}

Before proving Proposition~\ref{prop:inter_2_balls}, we introduce
some useful notations. For $0 \leq d \leq r$, we denote $U_d =
R({\bf I}_r | {\bf P}_d) \in E_r(q,n)$, where
${\bf P}_d = \left(\begin{array}{c|c} {\bf I}_d & {\bf 0} \\
\hline {\bf 0} & {\bf 0} \end{array}\right) \in \mathrm{GF}(q)^{r
\times (n-r)}$, hence $\di(U_0, U_d) = d$ for all $0 \leq d \leq r$.
We also denote the set of all generator matrices of all subspaces in
$B_u(U_0) \cap B_s(U_d)$ as $F(u,s,d)$, hence $|F(u,s,d)| =
\Ic(u,s,d) \prod_{i=0}^{r-1} (q^r-q^i)$.

\begin{lemma}\label{lemma:U0_Ud}
Let ${\bf X} = ({\bf A} | {\bf B}) \in \mathrm{GF}(q)^{r \times n}$,
where ${\bf A}$ and ${\bf B}$ have $r$ and $n-r$ columns,
respectively. Furthermore, we denote ${\bf A} = ({\bf A}_1 | {\bf a}
| {\bf A}_2)$ and ${\bf B} = ({\bf B}_1 | {\bf b} | {\bf B}_2)$,
where ${\bf a}$ and ${\bf b}$ are the $d$-th columns of ${\bf A}$
and ${\bf B}$, respectively. Then ${\bf X} \in F(u,s,d)$ if and only
if $\rk({\bf X}) = r$, $\rk({\bf B}) \leq u$, and $\rk({\bf B}_1 -
{\bf A}_1 | {\bf b} - {\bf a} | {\bf B}_2) \leq s$.
\end{lemma}

\begin{proof}
First, ${\bf X}$ is the generator matrix of some $V \in E_r(q,n)$ if
and only if $\rk({\bf X}) = r$. Also, it is easily shown that
$\di(V, U_0) = \rk({\bf B})$ and $\di(V, U_d) = \rk({\bf B} - {\bf
A} {\bf P}_d) = \rk({\bf B}_1 - {\bf A}_1 | {\bf b} - {\bf a} | {\bf
B}_2)$. Therefore, ${\bf X} \in F(u,s,d)$ if and only if $\rk({\bf
X}) = r$, $\rk({\bf B}) \leq u$, and $\rk({\bf B}_1 - {\bf A}_1 |
{\bf b} - {\bf a} | {\bf B}_2) \leq s$.
\end{proof}

We now give the proof of Proposition~\ref{prop:inter_2_balls}.

\begin{proof}
It suffices to show that $\Ic(u,s,d) \leq \Ic(u,s,d-1)$ for any $d
\geq 1$. We do so by first defining a mapping $\phi$ from $F(u,s,d)$
to $F(u,s,d-1)$ and then proving it is injective. Let ${\bf X} \in
F(u,s,d)$, then by Lemma \ref{lemma:U0_Ud}, $\rk({\bf X}) = r$,
$\rk({\bf B}) \leq u$, and $\rk({\bf B}_1 - {\bf A}_1 | {\bf b} -
{\bf a} | {\bf B}_2) \leq s$. Since the mapping $\phi$ only modifies
${\bf b}$, we shall denote $\phi({\bf X}) = {\bf Y} = ({\bf A} |
{\bf B}_1 | {\bf c} | {\bf B}_2)$. We hence have to show that
$\rk({\bf Y}) = r$, $\rk({\bf B}_1 | {\bf c} | {\bf B}_2) \leq u$,
and $\rk({\bf B}_1 - {\bf A}_1 | {\bf c} | {\bf B}_2) \leq s$. We
need to distinguish three cases.

\begin{itemize}
    \item Case I: $\rk({\bf B}_1 - {\bf A}_1 | {\bf B}_2)
    \leq s-1$. In this case, ${\bf c} = {\bf b}$. Note that
    $\rk({\bf Y}) = r$, $\rk({\bf B}) \leq u$, and
    $\rk({\bf B}_1 - {\bf A}_1 | {\bf c} | {\bf B}_2)
    \leq \rk({\bf B}_1 - {\bf A}_1 | {\bf B}_2) + 1 \leq s$.

    \item Case II: $\rk({\bf B}_1 - {\bf A}_1 | {\bf B}_2)
    = s$ and $\rk({\bf B}_1 | {\bf B}_2) \leq u-1$. In this
    case, ${\bf c} = {\bf b} - {\bf a}$. Note that $\rk({\bf Y}) =
    r$, $\rk({\bf B}_1 | {\bf c} | {\bf B}_2) \leq \rk({\bf B}) + 1 \leq u$, and
    $\rk({\bf B}_1 - {\bf A}_1 | {\bf c} | {\bf B}_2)
    = \rk({\bf B}_1 - {\bf A}_1 | {\bf b} - {\bf a} | {\bf B}_2)
    = s$.

    \item Case III: $\rk({\bf B}_1 - {\bf A}_1 | {\bf B}_2)
    = s$ and $\rk({\bf B}_1 | {\bf B}_2) = u$. We denote the column
    space of a matrix ${\bf D}$ as $C({\bf D})$. We have
    ${\bf b} - {\bf a} \in C({\bf B}_1 - {\bf A}_1 | {\bf
    B}_2)$ and ${\bf b} \in C({\bf B}_1 | {\bf
    B}_2)$. Hence ${\bf a} \in C({\bf B}_1 | {\bf
    B}_2 | {\bf B}_1 - {\bf A}_1)$. Denoting $C({\bf B}_1 | {\bf
    B}_2 | {\bf B}_1 - {\bf A}_1) = C({\bf B}_1 | {\bf B}_2)
    \oplus S$, where $S$ is a fixed subspace of
    $C({\bf B}_1 - {\bf A}_1)$,  ${\bf a}$ can be
    uniquely expressed as ${\bf a} = {\bf r} + {\bf s}$, where
    ${\bf r} \in C({\bf B}_1 | {\bf B}_2)$ and
    ${\bf s} \in S$.
    In this case, ${\bf c} = {\bf b} - {\bf r}$.
    Since ${\bf b} \in C({\bf B}_1 | {\bf B}_2)$,
    $\rk({\bf X}) = \rk({\bf A} | {\bf B}_1 | {\bf B}_2) = r = \rk({\bf Y})$.
    Also, since ${\bf c} \in C({\bf B}_1 | {\bf B}_2)$,
    $\rk({\bf B}_1 | {\bf c} | {\bf B}_2) = \rk({\bf B}_1 | {\bf B}_2) = u$.
    Finally, ${\bf c} = {\bf b} - {\bf a} + {\bf s}
    \in C({\bf B}_1 - {\bf A}_1 | {\bf B}_2)$,
    therefore
    $\rk({\bf B}_1 - {\bf A}_1 | {\bf c} | {\bf B}_2)
    = s$.
\end{itemize}

It is easy to show that $\phi$ is injective. Therefore, $|F(u,s,d)|
\leq |F(u,s,d-1)|$ and $\Ic(u,s,d) \leq \Ic(u,s,d-1)$.
\end{proof}

\subsection{Proof of
Proposition~\ref{prop:excess_bound}}\label{app:prop:excess_bound}

We adapt below the notations in \cite{vanwee_it88, vanwee_jct91} to
the injection metric for CDCs. For all $V \subseteq E_r(q,n)$ and a
CDC $\mathcal{C} \subseteq E_r(q,n)$ with covering radius $\rho$,
the excess on $V$ by $\mathcal{C}$ is defined to be
$E_\mathcal{C}(V) \df \sum_{C \in \mathcal{C}}|B_\rho(C) \cap V| -
|V|$. Hence if $\{W_i\}$ is a family of disjoint subsets of
$E_r(q,n)$, then $E_\mathcal{C} \left( \bigcup_i W_i \right) =
\sum_i E_\mathcal{C}(W_i)$. We define $\mathcal{Z} \df \{Z \in
E_r(q,n) : E_\mathcal{C}(\left\{ Z \right\}) \geq 1 \}$, i.e.,
$\mathcal{Z}$ is the set of subspaces covered by at least two
codewords in $\mathcal{C}$. It follows that $|\mathcal{Z}| \leq
E_\mathcal{C}(\mathcal{Z}) = E_\mathcal{C}(E_r(q,n)) = |\mathcal{C}|
\Vc(\rho) - {n \brack r}$.

Before proving Proposition \ref{prop:excess_bound}, we need the
following adaptation of \cite[Lemma 8]{vanwee_jct91}. Let
$\mathcal{C}$ be a code in $E_r(q,n)$ with covering radius $\rho$.
We define $\mathcal{A} \df \{U \in E_r(q,n) : \di(U,\mathcal{C}) =
\rho\}$.

\begin{lemma}\label{lemma:epsilon}
For $U \in \mathcal{A} \backslash \mathcal{Z}$ and $0 < \rho < r$,
we have $E_\mathcal{C}(B_1(U)) \geq \epsilon$.
\end{lemma}

\begin{proof}
Since $U \notin \mathcal{Z}$, there is a unique $C_0 \in
\mathcal{C}$ such that $\di(U, C_0) = \rho$. We have $|B_{\rho}(C_0)
\cap B_1(U)| = \Ic(\rho,1,\rho) = \Jc(\rho,0,\rho) +
\Jc(\rho,1,\rho) + \Jc(\rho-1,1,\rho) = 1 + a_\rho + c_\rho$. For
any codeword $C_1 \in \mathcal{C}$ satisfying $\di(U, C_1) = \rho +
1$, by Lemma~\ref{lemma:recursion_Jc} we have $|B_\rho(C_1) \cap
B_1(U)| = \Jc(\rho,1,\rho+1) = c_{\rho+1}$. Finally, for all other
codewords $C_2 \in \mathcal{C}$ at distance $> \rho+1$ from $U$, we
have $|B_\rho(C_2) \cap B_1(U)| = 0$. Denoting $N \df |\{ C_1 \in
\mathcal{C} : \di(U,C_1) = \rho+1 \}|$, we obtain
\begin{eqnarray*}
    E_\mathcal{C}(B_1(U))&=& \sum_{C \in \mathcal{C}}
    |B_{\rho}(C) \cap B_1(U)| - |B_1(U)|\\
    &=& 1+a_\rho+c_\rho + N c_{\rho+1} - \Nc(1) - 1=-b_\rho +N c_{\rho+1}\\
    &\equiv& -b_\rho \mod c_{\rho+1}.
\end{eqnarray*}
The proof is completed by realizing that $-b_\rho < 0$, while
$E_\mathcal{C}(B_1(U))$ is a non-negative integer.
\end{proof}

We now establish a key lemma.
\begin{lemma}\label{lemma:A_cap_B}
If $Z \in \mathcal{Z}$ and $0 < \rho < r$, then $|\mathcal{A} \cap
B_1(Z)| \leq \Vc(1) - c_\rho.$
\end{lemma}

\begin{proof}
By definition of $\rho$, there exists $C \in \mathcal{C}$ such that
$\di(Z,C) \leq \rho$. By Proposition~\ref{prop:inter_2_balls},
$|B_1(Z) \cap B_{\rho-1}(C)| \geq c_\rho$, with equality achieved
for $\di(Z,C) = \rho$. A subspace at distance $\leq \rho - 1$ from
any codeword does not belong to $\mathcal{A}$. Therefore, $B_1(Z)
\cap B_{\rho-1}(C) \subseteq B_1(Z) \backslash \mathcal{A}$, and
hence $|\mathcal{A} \cap B_1(Z)| = |B_1(Z)| - |B_1(Z) \backslash
\mathcal{A}| \leq \Vc(1) - |B_1(Z) \cap B_{\rho-1}(C)|$.
\end{proof}

We now give a proof of Proposition~\ref{prop:excess_bound}.

\begin{proof}
For a code $\mathcal{C}$ with covering radius $\rho$ and $\epsilon
\geq 1$,
\begin{eqnarray}
    \label{eq:gamma}
    \gamma & \df & \epsilon \left\{ {n \brack r} - |\mathcal{C}|\Vc(\rho-1)
    \right\}
    - (\epsilon-1) \left\{ |\mathcal{C}|\Vc(\rho) - {n \brack r}\right\} \\
    \label{eq:vanwee1}
    & \leq & \epsilon |\mathcal{A}| - (\epsilon-1)|\mathcal{Z}|\\
    \nonumber
    & \leq & \epsilon |\mathcal{A}| - (\epsilon-1) |\mathcal{A} \cap \mathcal{Z}|
    = \epsilon |\mathcal{A} \backslash \mathcal{Z}| + |\mathcal{A} \cap \mathcal{Z}|,
\end{eqnarray}
where~(\ref{eq:vanwee1}) follows from $|\mathcal{Z}| \leq
|\mathcal{C}|\Vc(\rho) - {n \brack r}$.

\begin{eqnarray}
    \label{eq:vanwee2}
    \gamma & \leq & \sum_{A \in \mathcal{A} \backslash \mathcal{Z}}
    E_\mathcal{C}(B_1(A)) + \sum_{A \in \mathcal{A} \cap \mathcal{Z}}
    E_\mathcal{C}(B_1(A))\\
    \nonumber
    & = & \sum_{A \in \mathcal{A}} E_\mathcal{C}(B_1(A)),
\end{eqnarray}
where~(\ref{eq:vanwee2}) follows from Lemma~\ref{lemma:epsilon} and
$|\mathcal{A} \cap \mathcal{Z}| \leq E_\mathcal{C}(\mathcal{A} \cap
\mathcal{Z})$.

\begin{eqnarray}
    \label{eq:vanwee3}
    \gamma & \leq & \sum_{A \in \mathcal{A}} \sum_{U \in B_1(A) \cap
    \mathcal{Z}} E_\mathcal{C}(\{ U \})\\
    \nonumber
    & = & \sum_{U \in \mathcal{Z}} \sum_{A \in B_1(U) \cap
    \mathcal{A}} E_\mathcal{C}(\{ U \})
    = \sum_{U \in \mathcal{Z}} |\mathcal{A} \cap B_1(U)| E_\mathcal{C}(\{ U
    \}),
\end{eqnarray}
where~(\ref{eq:vanwee3}) follows the fact that the second summation
is over disjoint sets $\{ U \}$. By Lemma~\ref{lemma:A_cap_B}, we
obtain
\begin{eqnarray}
    \nonumber
    \gamma & \leq & \sum_{U \in \mathcal{Z}} \left(\Vc(1) - c_\rho\right) E_\mathcal{C}(\{ U
    \})\\
    \nonumber
    & = & \left(\Vc(1) - c_\rho \right) E_\mathcal{C}(\mathcal{Z})\\
    \label{eq:vanwee4}
    & = & \left(\Vc(1) - c_\rho \right) \left\{ |\mathcal{C}|\Vc(\rho) - {n \brack r} \right\}.
\end{eqnarray}
Combining~(\ref{eq:vanwee4}) and~(\ref{eq:gamma}), we obtain the
bound in Proposition~\ref{prop:excess_bound}.
\end{proof}

\subsection{Proof of Lemma \ref{lemma:Kc(rho+1)}}
\label{app:lemma:Kc(rho+1)}

\begin{proof}
Let $\mathcal{C}$ be a code in $E_r(q,n-1)$ with covering radius
$\rho-1$ and cardinality $\Kc(q,n-1,r,\rho-1)$. Define the code
$\mathcal{C}_1 \subseteq E_r(q,n)$ as $\mathcal{C}_1 = \{R({\bf C} |
{\bf 0}) : R({\bf C}) \in \mathcal{C}\}$. For any $U_1 \in E_r(q,n)$
with generator matrix ${\bf U}_1 = ({\bf U} | {\bf u})$, where ${\bf
U} \in \mathrm{GF}(q)^{r \times n-1}$ and ${\bf u} \in
\mathrm{GF}(q)^{r \times 1}$, we prove that there exists $C_1 \in
\mathcal{C}_1$ generated by ${\bf C}_1 = ({\bf C} | {\bf 0})$ such
that $\di(C_1, U_1) \leq \rho$. We remark that $\rk({\bf U})$ is
equal to either $r$ or $r-1$. First, if $\rk({\bf U}) = r$, then
there exists $C \in \mathcal{C}$ such that $\rk({\bf C}^T | {\bf
U}^T) \leq r + \rho-1$. Second, if $\rk({\bf U}) = r-1$, then let
${\bf U}_0$ be $r-1$ linearly independent rows of ${\bf U}$. For any
${\bf v} \in \mathrm{GF}(q)^{n-1}$, ${\bf v} \notin R({\bf U}_0)$,
there exists $C \in \mathcal{C}$ such that $r + \rho-1 \geq \rk({\bf
C}^T | {\bf U}_0^T | {\bf v}^T) \geq \rk({\bf C}^T | {\bf U}_0^T) =
\rk({\bf C}^T | {\bf U}^T)$. Hence $\rk({\bf C}_1^T | {\bf U}_1^T)
\leq r + \rho$ and $\di(C_1, U_1) \leq \rho$. Thus $\mathcal{C}_1$
has covering radius at most $\rho$ and hence $\Kc(q,n,r,\rho) \leq
\Kc(q,n-1,r,\rho-1)$, which applied $\rho$ times yields
$\Kc(q,n,r,\rho) \leq \Kc(q,n-\rho,r,0) = {n-\rho \brack r}$.

Similarly, let $\mathcal{D}$ be a code in $E_{r-1}(q,n)$ with
covering radius $\rho-1$ and cardinality $\Kc(q,n,r-1,\rho-1)$.
Define the code $\mathcal{D}_1 = \{R(({\bf D}^T | {\bf d}^T)^T) :
R({\bf D}) \subseteq \mathcal{D}\} \in E_r(q,n)$, where ${\bf d} \in
\mathrm{GF}(q)^n$ is chosen at random such that $\rk({\bf D}^T |
{\bf d}^T) = r$. We remark that $|\mathcal{D}_1| \leq
|\mathcal{D}|$. For any $V_1 \in E_r(q,n)$ with generator matrix
${\bf V}_1 = ({\bf V}^T | {\bf v}^T)^T$, there exists $D_1 \in
\mathcal{D}_1$ with generator matrix ${\bf D}_1 = ({\bf D}^T | {\bf
d}^T)^T$ with $\rk({\bf D}^T | {\bf V}^T) \leq r + \rho - 2$. Thus
$\rk({\bf D}_1^T | {\bf V}_1^T) \leq r+\rho$ and $\mathcal{D}_1$ has
covering radius at most $\rho$. Thus $\Kc(q,n,r,\rho) \leq
|\mathcal{D}_1| \leq \Kc(q,n,r-1,\rho-1)$ which applied $\rho$ times
yields $\Kc(q,n,r,\rho) \leq \Kc(q,n,r-\rho,0) = {n \brack r-\rho}$.
\end{proof}

\subsection{Proof of Proposition \ref{prop:bound_combinatorial}}
\label{app:prop:bound_combinatorial}

\begin{proof}
Denoting the set of all codes of cardinality $K$ in $E_r(q,n)$ as
$S_K$, we have $|S_K| = {Q \choose K}$, where $Q \df {n \brack r}$.
For any code $\mathcal{C} \in K$ we denote the number of subspaces
in $E_r(q,n)$ at distance $> \rho$ from $\mathcal{C}$ as
$P(\mathcal{C})$.  The average value of $P(\mathcal{C})$ for all
codes $\mathcal{C} \in S_K$ is given by
\begin{eqnarray}
    \nonumber
    \frac{1}{|S_K|} \sum_{\mathcal{C} \in S_K} P(\mathcal{C})
    &=& \frac{1}{|S_K|}\sum_{\mathcal{C} \in S_K}
    \sum_{\stackrel{U \in E_r(q,n)}{\di(U, \mathcal{C}) > \rho}} 1\\
    \nonumber
    &=& \frac{1}{|S_K|}  \sum_{U \in E_r(q,n)}
    \sum_{\stackrel{\mathcal{C} \in S_K} {\di(U, \mathcal{C}) > \rho}} 1\\
    \label{eq:Q_choose_K}
    &=& \frac{1}{|S_K|} \sum_{U \in E_r(q,n)} {Q - \Vc(\rho) \choose K}\\
    \nonumber
    &=& \frac{Q}{|S_K|} {Q - \Vc(\rho) \choose K}.
\end{eqnarray}
Eq.~(\ref{eq:Q_choose_K}) comes from the fact that there are ${Q -
\Vc(\rho) \choose K}$ codes with cardinality $K$ that do not cover
$U$. For all $K$, there exists a code $\mathcal{C}' \in S_K$ for
which $P(\mathcal{C}')$ is no more than the average, i.e.,
$P(\mathcal{C}') \leq Q {Q \choose K}^{-1} {Q - \Vc(\rho) \choose K}
\leq Q \left( 1-Q^{-1}\Vc(\rho)\right)^K$. For $K = \left\lfloor
-\frac{1}{\log_Q \left(1-Q^{-1}\Vc(\rho) \right)} \right\rfloor +
1$, $P(\mathcal{C}') \leq Q \left(1- Q^{-1} \Vc(\rho) \right)^K < 1$
and $\mathcal{C}'$ has covering radius at most $\rho$.
\end{proof}

\subsection{Proof of Proposition~\ref{prop:bound_domination}}
\label{app:prop:bound_domination}

\begin{proof}
The proof is by induction on $k$. First, an augmented KK code is a
code with cardinality $k_0$ and minimum distance $2\rho+1$ for
$2\rho < r$, which hence leaves $u_{k_0}$ subspaces uncovered; for
$2 \rho \geq r$, a single codeword covers $\Vc(\rho)$ subspaces.
Second, suppose there exists a code with cardinality $k$ which
leaves \textbf{exactly} $v_k$ ($v_k \leq u_k$) subspaces uncovered,
and denote the set of uncovered subspaces as $V_k$. Let $G$ be the
graph where the vertex set is $E_r(q,n)$ and two vertices are
adjacent if and only if their distance is at most $\rho$. Let ${\bf
A}$ be the adjacency matrix of $G$ and ${\bf A}_k$ be the $v_k$
columns of ${\bf A}$ corresponding to $V_k$. There are $v_k
\Vc(\rho)$ ones in ${\bf A}_k$, distributed across $|N(V_k)|$ rows,
where $N(V_k)$ is the neighborhood \cite{godsil_book01} of $V_k$. By
construction, $N(V_k)$ does not contain any codeword, hence
$|N(V_k)| \leq {n \brack r} - k$. Also, by Lemma~\ref{lemma:B},
$|N(V_k)| \leq \Bc(v_k, \rho) \leq \Bc(u_k, \rho)$. Thus $|N(V_k)|
\leq \min \left\{ {n \brack r} - k, \Bc(u_k, \rho) \right\}$ and
there exists a row with at least $
    \left\lceil \frac{v_k \Vc(\rho)}
    {\min \left\{ {n \brack r} - k, \Bc(u_k, \rho) \right\}} \right\rceil
$ ones in ${\bf A}_k$. Adding the subspace corresponding to this row
to the code, we obtain a code with cardinality $k+1$ which leaves at
most $v_k - \left\lceil \frac{v_k \Vc(\rho)} {\min \left\{ {n \brack
r} - k, \Bc(u_k, \rho) \right\}} \right\rceil \leq u_{k+1}$
subspaces uncovered.
\end{proof}

\bibliographystyle{IEEETran}
\bibliography{gpt}

\end{document}